\documentclass[final,aps,prb,twocolumn,superscriptaddress,showpacs]{revtex4-1}

\usepackage{graphicx}
\usepackage{dcolumn}
\usepackage{amsmath,bm}
\usepackage{color}%
\usepackage{multirow}
\usepackage{array} 
\usepackage{esvect}
\usepackage{braket}
\usepackage{makecell}
\usepackage[utf8]{inputenc}
\usepackage{kotex}
\usepackage{xr}
\usepackage{soul,xcolor}
\setstcolor{red} 
\usepackage{booktabs}

\makeatletter
\newcommand*{\addFileDependency}[1]{
  \typeout{(#1)}
  \@addtofilelist{#1}
  \IfFileExists{#1}{}{\typeout{No file #1.}}
}
\makeatother

\newcommand*{\myexternaldocument}[1]{%
    \externaldocument{#1}%
    \addFileDependency{#1.tex}%
    \addFileDependency{#1.aux}%
}

\myexternaldocument{SI}  

\begin{document}

\title{Sapphire-Side hBN Formation by Ni--Cr-Mediated Growth in a Stacked Ni--Cr/Boron/Sapphire Configuration: Effect of Boron Supply}

\author{Donghoi Kim}
\affiliation{Department of Information Display, Kyung Hee University, Seoul 02447, Republic of Korea}

\author{Aelim Ha}
\affiliation{Department of Physics, University of Seoul, 163 Seoulsiripdae-ro, Dongdaemun-gu, Seoul 02504, Republic of Korea}
\affiliation{ Department of Materials Science and Engineering, Yonsei University, 50 Yonsei-ro, Seodaemun-gu, Seoul 03722, Republic of Korea}

\author{Soohyung Park}
\affiliation{Department of Physics, University of Seoul, 163 Seoulsiripdae-ro, Dongdaemun-gu, Seoul 02504, Republic of Korea}

\author{Young Duck Kim}
\affiliation{Department of Physics and Research Institute for Basic Sciences, Kyung Hee University, Seoul 02447, Republic of Korea}

\author{Chinkyo Kim}
\email[Corresponding author. E-mail:]{ckim@khu.ac.kr}
\affiliation{Department of Information Display, Kyung Hee University, Seoul 02447, Republic of Korea}
\affiliation{Department of Physics and Research Institute for Basic Sciences, Kyung Hee University, Seoul 02447, Republic of Korea}

\date{\today}
\begin{abstract}
Metal-mediated growth provides an effective route to high-quality hexagonal boron nitride (hBN), but hBN formation on the metal does not necessarily ensure hBN formation on an adjoining dielectric substrate. Here, we investigate this distinction using a vertically stacked Ni--Cr/boron/sapphire configuration operated at 1550~$^\circ$C under atmospheric pressure, with the relative boron supply varied at fixed Ni:Cr = 8:2. hBN is detected by Raman spectroscopy on the gas-exposed alloy surface throughout the investigated Alloy:B range, whereas sapphire-side hBN appears only at the higher relative boron supplies of Alloy:B = 24:1 and 18:1, as confirmed by both Raman spectroscopy and XRD. The hBN (002) reflection further narrows as the boron supply increases from 24:1 to 18:1, corresponding to an increase in the apparent c-axis coherent length from approximately 86 to 143~nm. Additional crystalline reflections also emerge only under these higher-boron conditions. These results identify boron supply as a key parameter governing the transition from alloy-side hBN formation to concurrent sapphire-side hBN growth. Once formed, sapphire-side hBN occurs within a substantially reconstructed and chemically modified Al/O-rich sapphire environment rather than as a uniform film on an intact substrate. To our knowledge, this study provides the first demonstration of successful sapphire-side hBN formation by Ni--Cr-mediated growth.
\end{abstract}

\maketitle

\section{Introduction}

Metal-mediated solution growth, often implemented using molten metal fluxes, is an established route to high-quality hexagonal boron nitride (hBN), because transition-metal melts can dissolve and transport B- and N-containing species and provide favorable conditions for crystallization. Ni--Cr, Fe--Cr, and Cu--Cr solvents have produced high-quality hBN crystals with large dimensions, with solvent composition, nitrogen solubility, nucleation, and mass transport playing central roles in the growth process.\cite{Kubota-CM-20-1661,Hoffman-JCG-393-114,Liu-CGD-17-4932,Li-JMCC-8-9931,Zhang-JCG-562-126074} Cr-containing alloys are particularly effective because Cr can enhance nitrogen incorporation and modify hBN growth kinetics, while also strongly influencing interfacial chemistry and adhesion.\cite{Hoffman-JCG-393-114,Liu-CGD-17-4932,Tai-JMST-217-128} Ni--Cr alloy composition can also influence hBN growth and the resulting hBN/metal interface,\cite{Tai-JMST-217-128} underscoring that the role of Cr extends beyond the bulk properties of the metal growth medium.

An important extension of metal-mediated hBN growth is to establish hBN on an adjoining substrate rather than restricting crystallization to the metal solvent itself. Kubota \textit{et al.} demonstrated atmospheric-pressure liquid-phase deposition of high-quality hBN onto sapphire using a Ni--Mo solvent,\cite{Kubota-Science-317-932} showing that hBN growth from a metal solution can extend to a dielectric substrate. More directly relevant to metal-mediated hBN formation on sapphire, Shi \textit{et al.} reported vapor--liquid--solid growth of relatively uniform multilayer hBN at the interface between molten Fe$_{82}$B$_{18}$ and sapphire at 1250~$^\circ$C.\cite{Shi-NC-11-849} In that system, the pre-alloyed Fe--B melt supplied boron, promoted N$_2$ activation, transported B--N-related species, and supported hBN nucleation and lateral growth at the liquid--solid interface. These studies establish that metal-mediated hBN formation can be extended to sapphire when an appropriate metal/substrate reaction environment is achieved.

Whether a comparable sapphire-side growth regime can be established with Ni--Cr is less straightforward. Although Ni--Cr is well established as an effective atmospheric-pressure solvent for hBN crystallization, an earlier comparative experiment disclosed in a patent reported a different outcome when sapphire was introduced into an hBN-forming Ni--Cr system. Under conditions where the Ni--Cr solvent successfully recrystallized hBN in the absence of sapphire, contact with sapphire was accompanied by reaction between Cr and the substrate and no hBN recrystallization on sapphire.\cite{Taniguchi-US7811909B2} This contrasting outcome demonstrates that the ability of Ni--Cr to support hBN crystallization does not automatically establish a viable growth environment at the adjoining sapphire surface. More generally, the known reactivity of Ni(Cr)/$\alpha$-Al$_2$O$_3$ interfaces at elevated temperatures\cite{Valenza-JECS-40-521} and vapor-phase studies of hBN growth on sapphire indicate that the interfacial state of sapphire can substantially modify the growth environment. Nitridation or thin AlN interlayers can alter the resulting hBN structure,\cite{Chubarov-CGD-12-3215,Page-PRM-3-064001} while aggressive high-temperature conditions can produce sapphire decomposition, oxygen depletion, roughening, and chemically modified hBN/sapphire interfaces.\cite{Yang-APEX-11-051002,Bansal-ACSAMI-13-54516} These observations point to the sapphire-side reaction environment as an independent factor in determining the outcome of metal-mediated hBN growth.

These previous results define the central problem addressed here. Ni--Cr can provide an effective medium for hBN crystallization, but hBN formation at the alloy does not necessarily imply that hBN will also form on adjoining sapphire. The sapphire-side process must additionally establish a local chemical and structural environment capable of sustaining hBN nucleation and crystallization. Because boron availability is a fundamental parameter in metal-mediated hBN growth, varying the relative boron supply provides a direct means of probing the transition between alloy-side hBN formation and sapphire-side hBN growth.

Here, we investigate this transition using a vertically stacked Ni--Cr/boron/sapphire configuration operated at 1550~$^\circ$C under atmospheric pressure, with the Ni:Cr composition fixed at 8:2 and the relative boron supply varied through the Alloy:B ratio. We compare hBN formation on the gas-exposed alloy surface with hBN formation on the sapphire side across the boron-supply series and then examine the local morphology and chemistry of the hBN-positive sapphire surface. The results show that hBN forms on the alloy throughout the investigated boron-supply range, whereas increasing the relative boron supply drives the onset of hBN formation on sapphire. The sapphire-side hBN that emerges under these conditions is accompanied by increased c-axis stacking coherence, additional crystalline-phase formation, and extensive chemical and morphological transformation of the sapphire surface.

\section{Experimental methods}

\subsection{High-temperature stacked growth of hBN on sapphire}

\begin{figure}
\includegraphics[width=1.0\columnwidth]{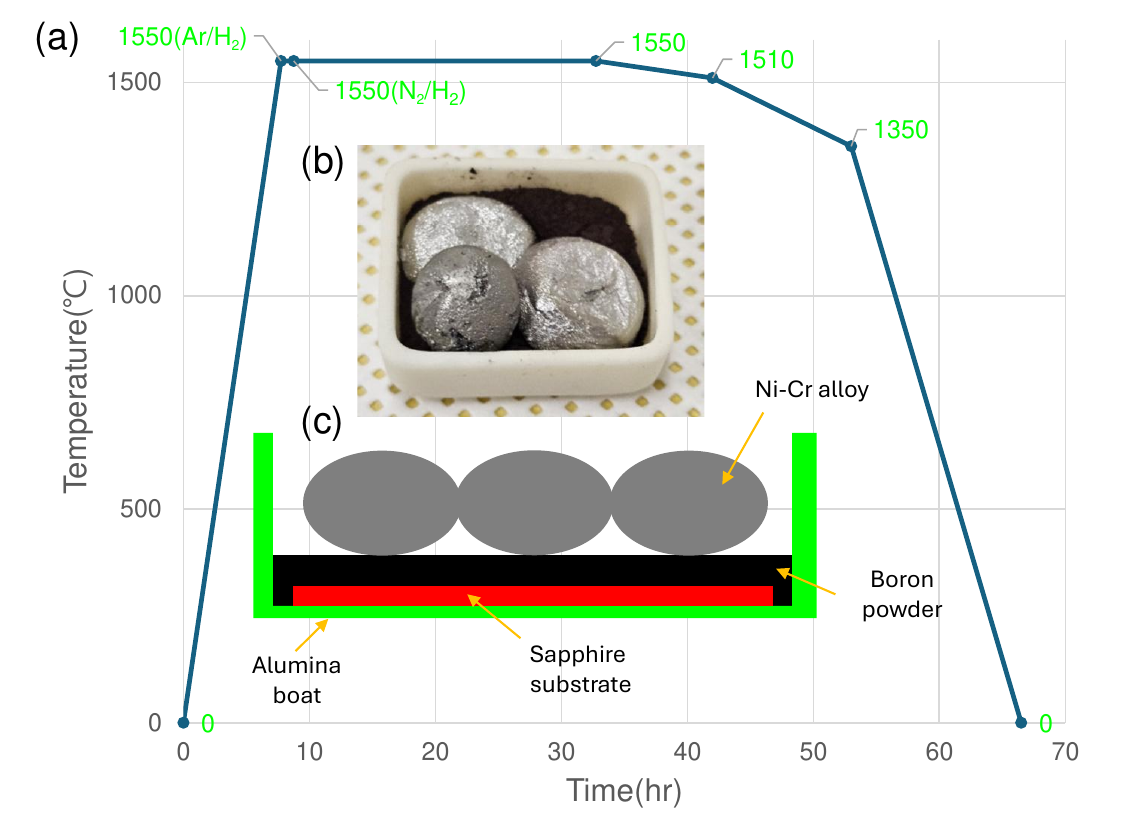}
\caption{Growth configuration and temperature profile used for hBN formation in the vertically stacked Ni--Cr/boron/sapphire geometry. (a) Temperature profile for the atmospheric-pressure high-temperature growth process. (b) Photograph of the alumina boat containing the Ni--Cr alloy, boron powder, and sapphire substrate before growth; the sapphire substrate is located beneath the boron powder and is therefore not visible. (c) Schematic illustration of the vertically stacked configuration, in which Ni--Cr alloy, boron powder, and sapphire substrate are arranged sequentially from top to bottom.}
\label{sample-configuration}
\end{figure}

A Ni--Cr alloy with a fixed Ni:Cr weight ratio of 8:2 was prepared from Ni and Cr powders with a total mass of 12~g. Ni powder (99.7\% trace metals basis, $<50~\mu\mathrm{m}$, Sigma-Aldrich) and Cr powder ($\geq$99\% trace metals basis, $<45~\mu\mathrm{m}$, Sigma-Aldrich) were mixed and heated in an alumina boat under flowing Ar/H$_2$ (250/50~sccm) at atmospheric pressure from room temperature to 1550~$^\circ$C at 200~$^\circ$C~h$^{-1}$, held for 24~h, and cooled to room temperature at 100~$^\circ$C~h$^{-1}$.

For hBN growth, a $c$-plane sapphire substrate, amorphous boron powder (99\%, 2N, MSE Supplies), and the preformed Ni--Cr alloy were vertically stacked from bottom to top in an alumina boat, as shown in Fig.~\ref{sample-configuration}(c). The assembly was heated to 1550~$^\circ$C at 200~$^\circ$C~h$^{-1}$ under Ar/H$_2$ (250/50~sccm), held for 1~h, and then exposed to N$_2$/H$_2$ (125/5~sccm) for 24~h. The sample was subsequently cooled from 1550 to 1510~$^\circ$C at approximately 4.3~$^\circ$C~h$^{-1}$, from 1510 to 1350~$^\circ$C at approximately 14.5~$^\circ$C~h$^{-1}$, and then to room temperature at 100~$^\circ$C~h$^{-1}$.

The relative boron supply was varied by changing the Alloy:B weight ratio to 36:1, 30:1, 24:1, and 18:1 while maintaining the same Ni:Cr composition and growth conditions. The Alloy:B weight ratio is used here as an operational measure of the relative boron supply. After growth, no residual boron powder was macroscopically observed between the Ni--Cr alloy and the sapphire substrate, and the alloy was firmly adhered to the sapphire and could not be separated mechanically. The gas-exposed alloy surface was characterized before removal of the alloy. The adhered alloy was then removed by wet etching in aqua regia (HCl:HNO$_3$ = 4:1) at 80~$^\circ$C for 30~min, and the exposed sapphire surface corresponding to the former alloy-contact region was subsequently characterized by Raman spectroscopy, XRD, SEM, EDS, and XPS.

\subsection{Structural, spectroscopic, and microscopic characterization}

Raman spectroscopy was used to identify hBN from the characteristic $E_{2g}$ mode near 1366--1367~cm$^{-1}$. Raman measurements were performed using a Renishaw inVia Raman spectrometer equipped with a 532~nm laser with a maximum output power of 50~mW. The laser power was set to 27.4~mW for the sapphire-side measurements and 0.19~mW for the alloy-side measurements. Spectra were acquired with an acquisition time of 1~s. Raman mapping was performed with step sizes of 0.5~$\mu$m on the sapphire side and 1~$\mu$m on the alloy side. Raman maps were constructed from the intensity near 1366~cm$^{-1}$ and correlated with optical and SEM images where applicable.

Optical microscopy was performed in reflection mode to identify and register regions for subsequent Raman and SEM measurements. SEM was used to examine the sapphire-side morphology in plan-view and tilted configurations, and a thin Pt coating was applied where necessary to reduce charging. Corresponding regions were identified across optical, Raman, and SEM images using characteristic surface features. SEM measurements were performed using a Hitachi S-4700 microscope at an accelerating voltage of 10~kV.

X-ray diffraction (XRD) measurements were performed using an X'Pert PRO diffractometer (PANalytical) equipped with a hybrid monochromator (2$\times$Ge(220), Cu asymmetric configuration). Cu K$\alpha_1$ radiation was used with the X-ray source operated at 45~kV and 30~mA.  The hBN (002) reflection near 26.5--26.6$^\circ$ was used to identify c-axis-stacked hBN.  The hBN (002) reflections of the Alloy:B = 18:1 and 24:1 samples were fitted using pseudo-Voigt line shapes to determine their full widths at half maximum (FWHM). The apparent coherent length along the c-axis was estimated using the Debye--Scherrer relation, $L_c=K\lambda/(\beta\cos\theta)$, where $K=0.9$, $\lambda=1.5406$~\AA\ for Cu K$\alpha_1$ radiation, $\beta$ is the fitted FWHM in radians, and $\theta$ is the Bragg angle of the hBN (002) reflection. Because instrumental broadening was not independently corrected, the resulting values are interpreted as apparent c-axis coherent lengths rather than direct measurements of the physical hBN thickness.

XPS measurements (Nexsa, Thermo Fisher Scientific) were performed to investigate the chemical states, utilizing a monochromatic Al K$\alpha$ source (h$\nu$ = 1486.68~eV).  Surface charging was compensated using a dual-beam charge neutralizer, and the binding energies were calibrated by referencing the adventitious C~1s peak to 284.6~eV.

\section{Results and discussion}

\subsection{Contrasting boron-supply dependence of hBN formation on the alloy and sapphire}

\begin{figure*}
\includegraphics[width=2.0\columnwidth]{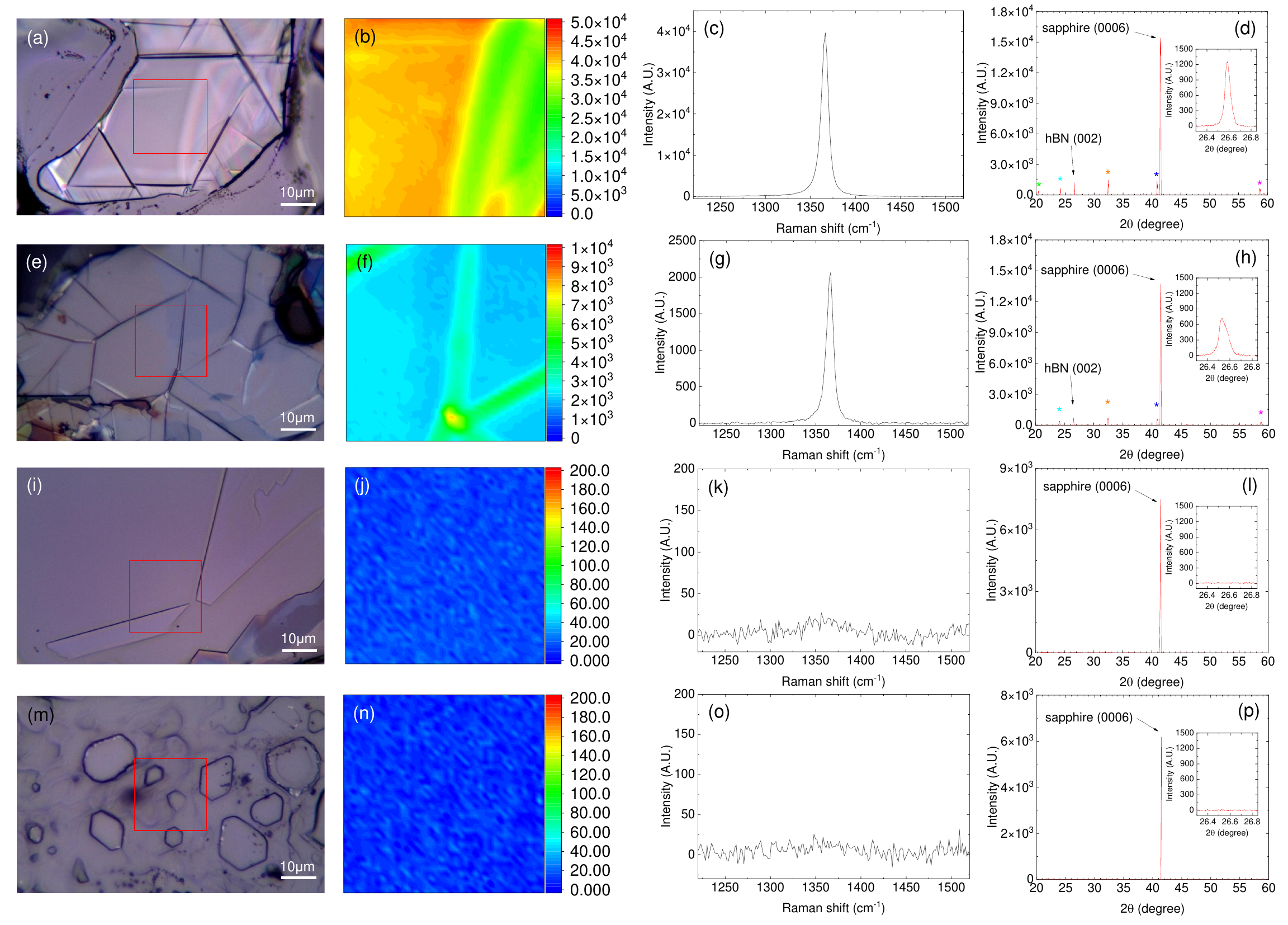}
\caption{Boron-supply-dependent characterization of the sapphire side after growth in the vertically stacked Ni--Cr/boron/sapphire configuration. (a)--(d), (e)--(h), (i)--(l), and (m)--(p) show optical microscopy images, Raman intensity maps, representative Raman spectra, and X-ray diffraction patterns for Alloy:B = 18:1, 24:1, 30:1, and 36:1, respectively. Red boxes in the optical images indicate the Raman-mapping regions. The maps were constructed from the intensity near the hBN $E_{2g}$ mode at 1366--1367~cm$^{-1}$, and the corresponding color scales represent Raman intensity in arbitrary units (a.u.), with blue and red indicating lower and higher intensities, respectively.  The laser power was set to 27.4~mW. Consistently, the XRD patterns for the 18:1 and 24:1 samples exhibit an hBN (002) peak near 26.5--26.6$^\circ$, while this peak is not discernible for the 30:1 and 36:1 samples.  The peak near 41.4$^\circ$ is assigned to sapphire (0006), and asterisks denote unidentified secondary crystalline reflections.  Insets in (d), (h), (l), and (p) show enlarged views of the hBN (002) region.}
\label{Boron-dependence}
\end{figure*}

Figure~\ref{Boron-dependence} establishes a clear boron-supply dependence of hBN formation on the sapphire side at 1550~$^\circ$C. For Alloy:B = 18:1, optically distinct regions containing pronounced straight and polygonal line features are observed on the sapphire surface [Fig.~\ref{Boron-dependence}(a)]. The corresponding Raman map exhibits a strong spatially localized response near the characteristic hBN $E_{2g}$ mode at approximately 1366~cm$^{-1}$ [Fig.~\ref{Boron-dependence}(b)], and the representative spectrum in Fig.~\ref{Boron-dependence}(c) shows a sharp peak at the same position. The XRD pattern in Fig.~\ref{Boron-dependence}(d) independently exhibits a clear hBN (002) reflection near 26.6$^\circ$. The 24:1 sample likewise shows both a detectable hBN-related Raman response and an hBN (002) diffraction peak [Fig.~\ref{Boron-dependence}(e)--(h)], although both responses are weaker than those observed for the 18:1 sample. By contrast, neither a distinct hBN Raman signature nor an hBN (002) reflection is detected for the 30:1 and 36:1 samples [Fig.~\ref{Boron-dependence}(i)--(p)], despite substantial sapphire-side morphological restructuring. The boron-supply series therefore shows the onset of detectable sapphire-side hBN between the 30:1 and 24:1 conditions, with hBN detected at the higher relative boron supplies of 24:1 and 18:1.

The enlarged XRD regions shown in the insets of Fig.~\ref{Boron-dependence}(d), (h), (l), and (p) further resolve the hBN (002) region. Pseudo-Voigt fitting of the hBN (002) reflection gives FWHM values of approximately 0.057$^\circ$ and 0.095$^\circ$ for the 18:1 and 24:1 samples, respectively. Using the Debye--Scherrer relation, these widths correspond to apparent c-axis coherent lengths of approximately 143 and 86~nm, respectively. The narrower reflection and larger apparent coherent length for the 18:1 sample indicate that increased boron supply is accompanied by enhanced out-of-plane stacking coherence. The extracted lengths represent coherent diffraction lengths along the c-axis rather than direct measurements of the physical BN thickness. In addition to the hBN (002) reflection and the intense sapphire (0006) substrate peak near 41.4$^\circ$, the 18:1 and 24:1 patterns contain additional peaks near 24.1$^\circ$, 32.4$^\circ$, 40.9$^\circ$, and 58.7$^\circ$. These reflections are absent from the 30:1 and 36:1 patterns and therefore emerge selectively under the higher-boron conditions. Their correlated appearance with sapphire-side hBN indicates that increased boron supply modifies the substrate-side reaction environment beyond hBN formation alone. Because their phase identity cannot be uniquely established from the present diffraction patterns, the peaks are denoted by asterisks and assigned to an unidentified boron-supply-dependent secondary crystalline phase.

\begin{figure*}
\includegraphics[width=2.0\columnwidth]{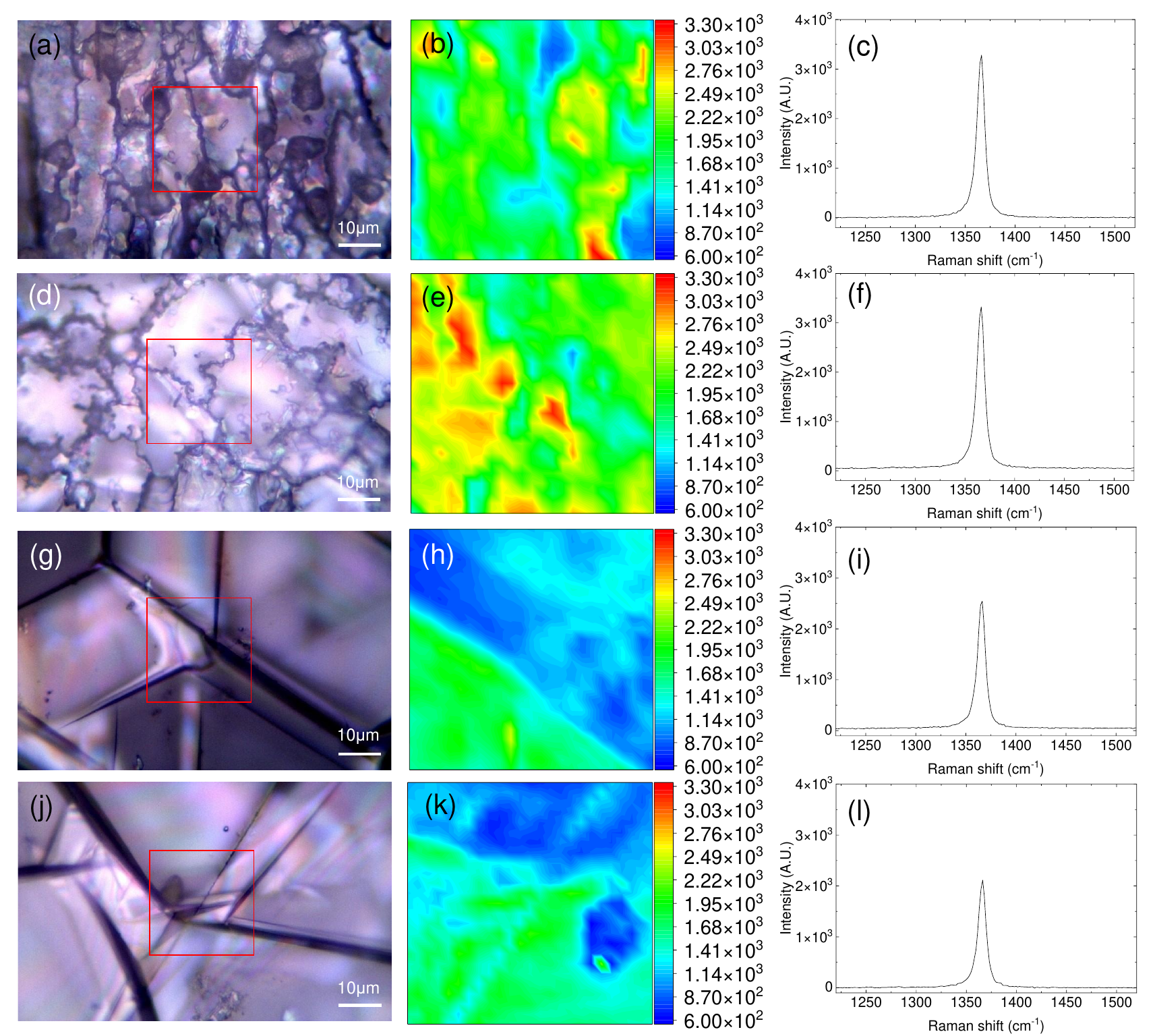}
\caption{Optical and Raman characterization of the gas-exposed Ni--Cr alloy surfaces after growth at fixed Ni:Cr = 8:2 and different Alloy:B weight ratios. (a)--(c) Alloy:B = 18:1, (d)--(f) 24:1, (g)--(i) 30:1, and (j)--(l) 36:1. (a), (d), (g), and (j) Optical microscopy images of the post-growth alloy surfaces, with the red boxes indicating the regions used for Raman mapping. (b), (e), (h), and (k) Raman intensity maps at 1366~cm$^{-1}$ obtained from the corresponding marked regions. (c), (f), (i), and (l) Representative Raman spectra showing the characteristic hBN $E_{2g}$ band near 1366--1367~cm$^{-1}$. hBN Raman signatures are observed on the gas-exposed alloy surface throughout the investigated Alloy:B range. The laser power was set to 0.19~mW.}
\label{Raman-on-alloy}
\end{figure*}

The sapphire-side behavior becomes particularly informative when compared with hBN formation on the gas-exposed Ni--Cr alloy surface. As shown in Fig.~\ref{Raman-on-alloy}, Raman mapping reveals the characteristic hBN response near 1366--1367~cm$^{-1}$ across the mapped alloy regions for all four Alloy:B ratios.  Raman point measurements performed at additional locations on each alloy surface consistently showed the same spectral characteristics, indicating that hBN is widespread over the gas-exposed alloy surface throughout the investigated boron-supply range. In contrast, sapphire-side hBN is detected only for Alloy:B = 24:1 and 18:1 and occupies approximately 10\% of the examined area within the sapphire region exposed after removal of the adhered Ni--Cr alloy. Thus, hBN formation on the alloy persists over a substantially wider boron-supply range than hBN formation on sapphire. Because the alloy- and sapphire-side Raman measurements were acquired at different laser powers, absolute Raman intensities are not compared quantitatively between the two surfaces.

Combining the alloy-side and sapphire-side measurements reveals two distinct boron-supply regimes. At Alloy:B = 36:1 and 30:1, hBN forms on the gas-exposed Ni--Cr alloy while no hBN is detected on the sapphire side. At Alloy:B = 24:1 and 18:1, hBN additionally emerges on sapphire. This transition demonstrates that hBN formation on the alloy and on sapphire has different effective growth requirements, with sapphire-side hBN requiring a higher relative boron supply. The XRD results further show that increasing boron supply affects not only the occurrence of sapphire-side hBN but also its crystalline state: the hBN (002) reflection narrows from 24:1 to 18:1, corresponding to an increase in the apparent c-axis coherent length from approximately 86 to 143~nm. In parallel, additional crystalline reflections appear only for the 24:1 and 18:1 conditions. Boron supply therefore controls the transition from alloy-side hBN formation to a distinct growth regime in which hBN additionally develops on sapphire with increased stacking coherence and accompanying substrate-side crystalline-phase formation.

The contrasting alloy-side and sapphire-side responses suggest that the role of boron supply is mediated through the Ni--Cr growth medium rather than through direct nitridation of boron at the sapphire surface. hBN forms on the gas-exposed alloy throughout the investigated Alloy:B range, indicating that the Ni--Cr alloy can support hBN crystallization even at relatively low boron supply. Sapphire-side hBN, however, appears only at Alloy:B = 24:1 and 18:1, suggesting that a higher B loading of the alloy or a higher local B--N supersaturation is required for transport and nucleation at the adjoining sapphire interface. In this picture, the Ni--Cr alloy serves not only as a site for hBN formation but also as a medium for dissolving and transporting B- and N-containing species toward the substrate-side interface. The concurrent sapphire reconstruction and Cr-related chemical signatures further indicate that hBN nucleation occurs within a chemically evolving metal--sapphire reaction environment. Thus, increased boron supply likely shifts the balance toward interfacial hBN precipitation while Cr-associated substrate reconstruction proceeds simultaneously.

This framework also provides a possible explanation for the contrasting result reported by Taniguchi \textit{et al.}\cite{Taniguchi-US7811909B2} In that study, pre-existing hBN was used as the BN source material and dissolved in a Cr-rich Ni--Cr solvent before recrystallization, and the Ni--Cr/sapphire experiment was performed at a single hBN-to-metal ratio. The effect of increasing boron availability within the Ni--Cr/sapphire system was therefore not examined. In contrast, the present system supplies elemental boron separately from continuously flowing N$_2$/H$_2$ and systematically varies the relative boron supply at fixed Ni:Cr composition. The observation that hBN forms on the alloy throughout the investigated Alloy:B range but appears on sapphire only at Alloy:B = 24:1 and 18:1 indicates that the substrate-side process requires a higher effective boron availability than alloy-side crystallization. Moreover, the XPS and Cr$^{3+}$ luminescence results, as examined below in subsection~\ref{XPS-PL}, show that Cr--sapphire interaction still occurs, so the successful formation of sapphire-side hBN cannot be attributed simply to suppression of this reaction. Rather, the present results are consistent with sufficiently high boron availability in the Ni--Cr-mediated growth environment allowing hBN nucleation at the reacting sapphire interface even while Cr-associated substrate reconstruction proceeds.

\subsection{Three-dimensional morphology of the hBN-positive 18:1 sapphire surface}

\begin{figure*}
\includegraphics[width=2.0\columnwidth]{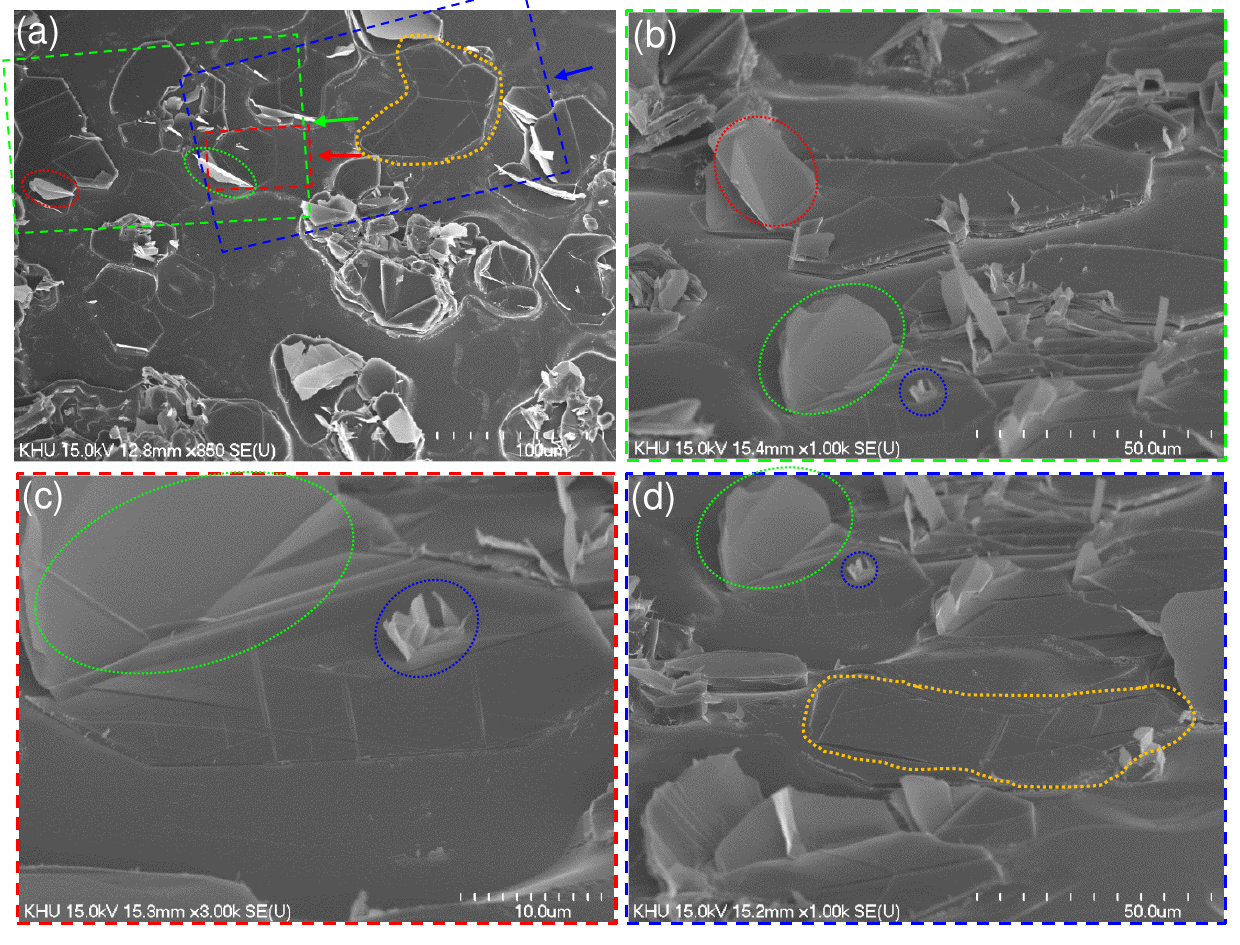}
\caption{Three-dimensional surface morphology of the Alloy:B = 18:1 sample. (a) Low-magnification SEM image showing the overall sapphire-side reconstructed morphology. The color-coded dashed boxes indicate the regions examined at higher magnification and from different viewing directions in (b)--(d), and the arrows indicate the corresponding viewing directions. Color-coded dotted outlines serve only as visual guides to identify the same objects across panels and do not represent distinct morphological classes. The orange-outlined region corresponds to the area examined in Fig.~\ref{Boron-dependence}(a).  (b)--(d) Higher-magnification SEM views revealing the faceted, stepped, and three-dimensional character of the reconstructed surface.}
\label{Morphology-landscape}
\end{figure*}

To place the locally hBN-positive morphology in a broader spatial context, the sapphire side of the 18:1 sample was examined over a substantially larger field of view by SEM, as shown in Fig.~\ref{Morphology-landscape}. Within the relatively limited region examined in Fig.~\ref{Boron-dependence}(a), the surface appears comparatively smooth and could be interpreted as a locally film-like hBN-bearing region. The wider SEM view in Fig.~\ref{Morphology-landscape}(a), however, shows that this morphology occupies only part of the overall sapphire-side surface. Much of the surrounding area consists of large reconstructed structures with pronounced three-dimensional relief, together with locally lifted or detached sheet-like features. The color-coded regions in (a) are shown at higher magnification and from different viewing directions in Fig.~\ref{Morphology-landscape}(b)--(d), where well-defined lateral boundaries, inclined facets, stepped or terrace-like relief, and partially detached sheet-like material are clearly resolved.

The orange-outlined region in Fig.~\ref{Morphology-landscape}(a) corresponds to the area examined in Fig.~\ref{Boron-dependence}(a), placing the locally hBN-positive region within this broader reconstructed landscape. The comparison demonstrates that sapphire-side hBN formation at 18:1 occurs within a strongly heterogeneous three-dimensional environment rather than on a uniformly preserved sapphire surface. Relatively smooth hBN-positive areas coexist with extensively reconstructed substrate regions and locally detached sheet-like material. The 18:1 growth outcome is therefore better described as hBN formation within a reconstructed sapphire environment than as conventional film growth on a flat substrate. The spatial relationship between Raman-positive hBN and this reconstructed morphology is examined below using correlated optical, Raman, SEM, and EDS measurements.

\subsection{Correlated optical, Raman, and SEM identification of local BN--sapphire configurations}

\begin{figure}
\includegraphics[width=1.0\columnwidth]{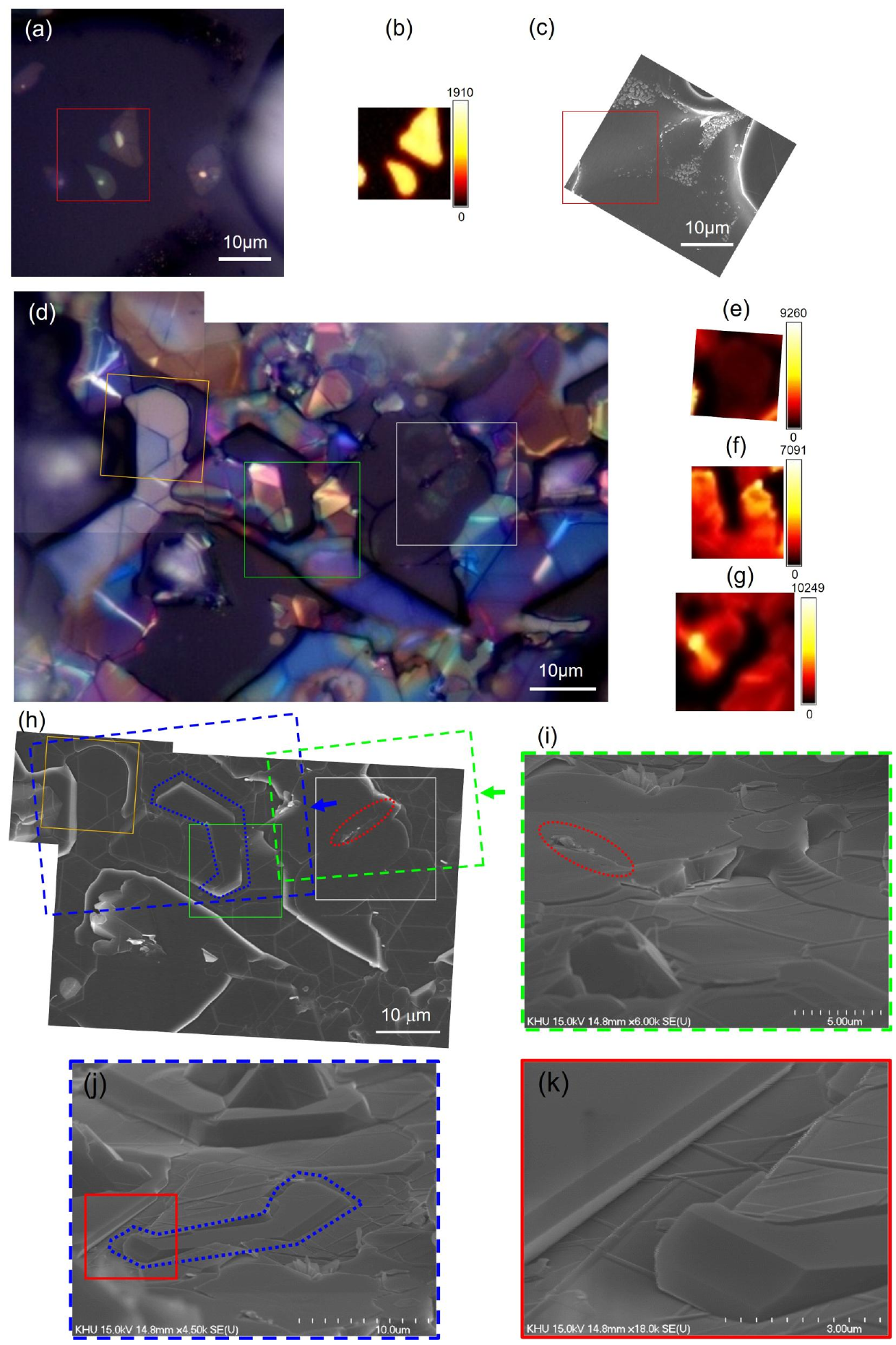}
\caption{Optical microscopy, Raman mapping, and SEM characterization of the Alloy:B = 18:1 sample. (a) Optical microscopy image and (b) corresponding Raman intensity map at 1366~cm$^{-1}$. (c) SEM image of the same region. (d) Optical microscopy image from a different area of the same sample, with (e)--(g) Raman intensity maps for the marked regions. The color scales in (b) and (e)--(g) represent Raman intensity in arbitrary units (a.u.). (h) SEM image of the region shown in (d). Color-coded dashed boxes, dotted outlines, and arrows guide identification of the corresponding objects and viewing directions in the tilted SEM images in (i) and (j). In (j), the Raman-active region corresponding to (g) does not exhibit an obvious separately resolved BN-specific surface morphology, supporting its Type II assignment. (k) Higher-magnification SEM image of the area marked by the red box in (j), illustrating the local relationship between linear surface features and the reconstructed morphology.}
\label{18-to-1-sample}
\end{figure}

Figure~\ref{18-to-1-sample} correlates optical contrast, Raman response, and SEM morphology within the strongly hBN-positive 18:1 sample. In the first region, the optically distinct domains visible in Fig.~\ref{18-to-1-sample}(a) coincide with strong Raman intensity near 1366~cm$^{-1}$ [Fig.~\ref{18-to-1-sample}(b)], identifying these regions as hBN-positive. In the corresponding SEM image [Fig.~\ref{18-to-1-sample}(c)], however, the Raman-positive domains do not appear as independently resolved hBN-specific structures at the outermost surface. The combined measurements therefore show that hBN can be present within the reconstructed region without producing a separately distinguishable surface morphology in SEM.

The larger region in Fig.~\ref{18-to-1-sample}(d) allows the observed behavior to be classified into three local configurations. Type I corresponds to reconstruction-dominated regions without a detectable hBN Raman response. Type II corresponds to Raman-positive regions in which hBN does not appear as a separately resolved outermost-surface morphology. Type III corresponds to Raman-positive regions accompanied by a distinct surface or near-surface morphological expression. The Raman maps in Fig.~\ref{18-to-1-sample}(e)--(g) identify Type II and Type III regions as hBN-positive, whereas Type I regions remain Raman-negative. In the corresponding SEM image [Fig.~\ref{18-to-1-sample}(h)], Type II and Type III therefore represent two different morphological manifestations of the same spectroscopically identified hBN-positive state.

The tilted SEM views in Fig.~\ref{18-to-1-sample}(i) and (j) provide additional three-dimensional context for this classification. The color-coded outlines register corresponding physical objects between the plan-view and tilted images. In particular, Fig.~\ref{18-to-1-sample}(j) examines the region corresponding to the Raman-active area mapped in Fig.~\ref{18-to-1-sample}(g). Despite its clear Raman response, this region lacks an independently resolved BN-specific outermost-surface feature, directly supporting its Type II assignment.   The absence of a separately resolved BN-specific surface morphology is therefore consistent with BN being located beneath or within the reconstructed surface region, although the present measurements do not establish a unique vertical location. The higher-magnification image in Fig.~\ref{18-to-1-sample}(k) further illustrates the close spatial relationship between linear surface features and the surrounding reconstructed morphology. These linear features are treated here as morphological markers of the reconstructed region rather than as a separately identified hBN morphology.

\begin{figure}
\includegraphics[width=1.0\columnwidth]{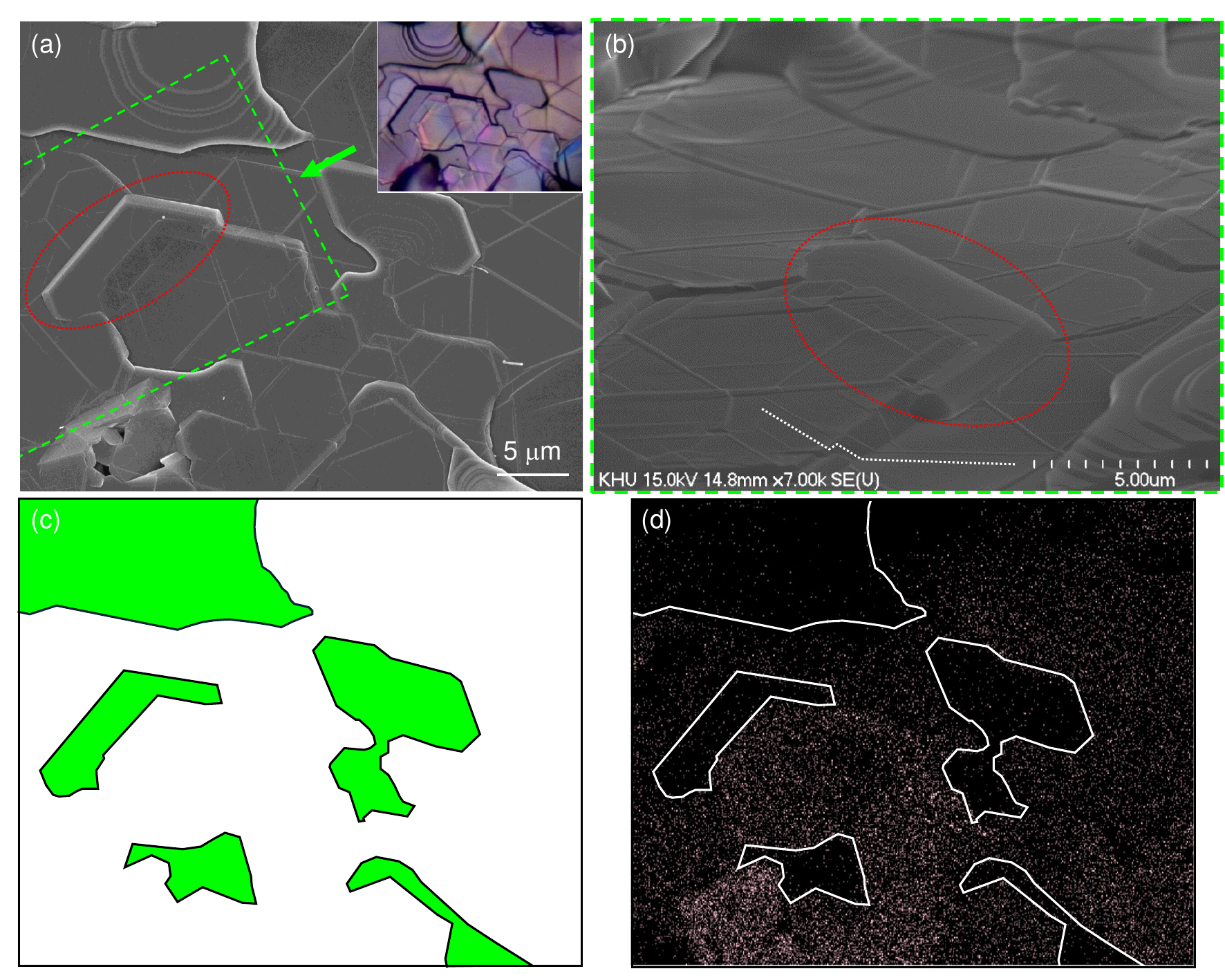}
\caption{EDS analysis and surface morphology of the Alloy:B = 18:1 sample. (a) SEM image of a different area from the same sample shown in Fig.~\ref{18-to-1-sample}; the inset shows the corresponding optical microscopy image. The green dashed outline marks the region examined at a tilted viewing angle in (b), and the green arrow indicates the viewing direction. (b) Tilted SEM image revealing the stepped and faceted three-dimensional morphology of the reconstructed sapphire surface. The white dotted lines serve only as guides to the eye for tracing corresponding linear surface features, and the red dotted outline identifies the same object in (a) and (b). (c) Type I regions highlighted in green. (d) N elemental map obtained by EDS, overlaid with the Type I outlines. The absence of clear N enhancement associated with the Type I regions supports their assignment as reconstruction-dominated regions rather than hBN-bearing surface states.}
\label{EDS}
\end{figure}

The EDS measurements in Fig.~\ref{EDS} provide an additional compositional distinction between the reconstruction-dominated and hBN-positive regions. The tilted SEM image in Fig.~\ref{EDS}(b) further illustrates the faceted and stepped three-dimensional character of the reconstructed surface. The Type I areas identified in Fig.~\ref{EDS}(c) do not exhibit corresponding localized enhancement in the N elemental map in Fig.~\ref{EDS}(d). Although EDS is not highly sensitive to very thin BN layers, the combined Raman and EDS results support the assignment of Type I as reconstruction-dominated regions without detectable hBN.

Taken together, the correlated optical, Raman, SEM, and EDS measurements identify three principal local outcomes within a single 18:1 specimen: reconstruction without detectable hBN (Type I), hBN incorporated into a region without a separately resolved hBN-specific surface morphology (Type II), and hBN accompanied by a distinct surface or near-surface morphological expression (Type III). Their coexistence demonstrates that entering the sapphire-side hBN growth regime does not produce a single uniform hBN/sapphire architecture. Instead, hBN formation proceeds within a spatially heterogeneous substrate environment whose local structural manifestation varies across the specimen.

This spatial heterogeneity adds a second level to the boron-supply dependence established above. Increased boron supply enables sapphire-side hBN formation, but the resulting hBN develops within an already strongly reconstructed and chemically evolving sapphire environment. The 18:1 condition most effectively supports sapphire-side hBN crystallization, while also exhibiting substantial substrate-side transformation.

\subsection{Surface chemistry of the reconstructed sapphire region}
\label{XPS-PL}

\begin{figure}
\includegraphics[width=1.0\columnwidth]{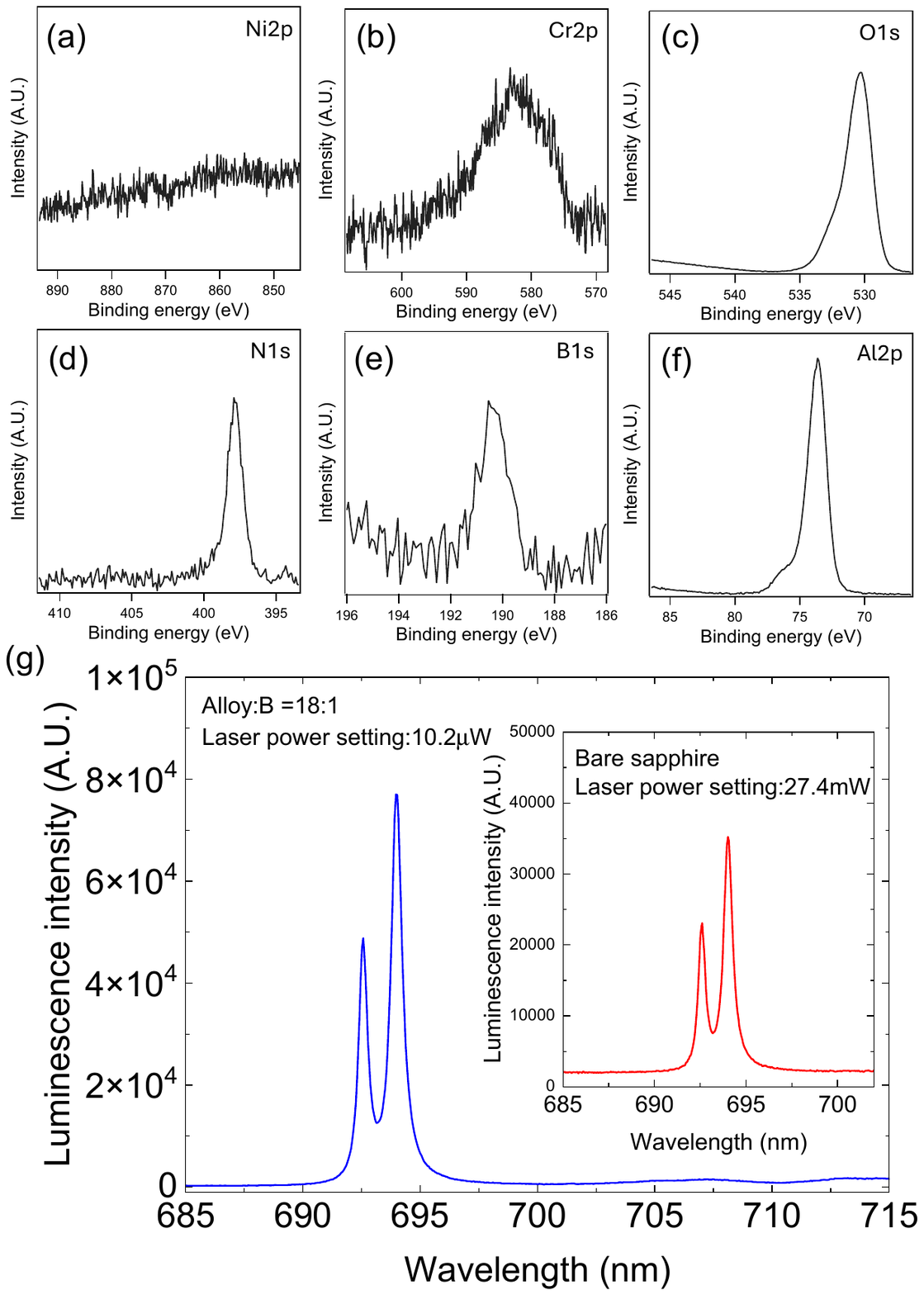}
\caption{XPS and Cr$^{3+}$-related luminescence characterization of the reconstructed sapphire-side region of the Alloy:B = 18:1 sample. (a)--(f) High-resolution XPS spectra of Ni 2p, Cr 2p, O 1s, N 1s, B 1s, and Al 2p, respectively. (g) Luminescence spectra acquired using the Raman microscope from the reconstructed Alloy:B = 18:1 sapphire-side region (blue) and bare sapphire (red, inset). The 18:1 sample and bare sapphire were measured at laser-power settings of 10.2$\mu$W and 27.4mW, respectively.}
\label{XPS}
\end{figure}

XPS was used to determine the chemical character of the reconstructed sapphire-side region and to distinguish substrate-derived reconstruction from residual alloy-derived material. As shown in Fig.~\ref{XPS}, the analyzed region is dominated by strong Al 2p and O 1s responses, demonstrating that the outer reconstructed surface remains primarily Al/O-rich. No distinct Ni 2p feature is resolved in either the metallic-Ni region near 852.6~eV or the Ni$^{2+}$ region near 854--856~eV. The Cr 2p response is much weaker and appears as a broad high-binding-energy feature near 581--582~eV, consistent with Cr being present predominantly in an oxidized rather than metallic environment. These results indicate that the reconstructed outer surface is principally substrate-derived, with a weak oxidized-Cr contribution and no clear XPS evidence for retained Ni.

Additional information on the local Cr environment is provided by the luminescence spectra in Fig.~\ref{XPS}(g). Both the reconstructed Alloy:B = 18:1 sapphire-side region and bare sapphire exhibit a doublet near 693--694~nm, consistent with the characteristic R$_1$/R$_2$ emission of Cr$^{3+}$ in an Al$_2$O$_3$-like local environment.\cite{Toyoda-MSEB-54-33,Michaels-MD-107-478} The 18:1 spectrum remains pronounced at a laser-power setting of only 10.2~$\mu$W, whereas the bare-sapphire spectrum was acquired at 27.4~mW. Because the spectra were measured at markedly different excitation-power settings, their raw intensities are not compared quantitatively. Nevertheless, the strong Cr$^{3+}$-related luminescence observed from the grown sample, together with the oxidized-Cr XPS response, supports the presence of Cr within an Al$_2$O$_3$-like environment in the reconstructed sapphire-side region.

The B 1s and N 1s spectra provide complementary evidence for BN-related bonding within the same reconstructed surface environment. The B 1s feature lies near 190.5--191.1~eV and the N 1s feature near 398.0--398.7~eV, within the expected ranges for B--N bonding. Their concurrent detection, together with the Raman and XRD identification of hBN, supports the presence of BN-related material within the reconstructed Al/O-rich sapphire-side region.

Taken together, the XPS and luminescence measurements show that the reconstructed surface is predominantly Al/O-rich, contains a weak oxidized-Cr component associated with an Al$_2$O$_3$-like local environment, and shows no clear evidence for a retained Ni-rich surface phase. The sapphire-side growth outcome is therefore more appropriately described as hBN formation within a chemically and morphologically transformed sapphire environment than as deposition of a uniform hBN film on an intact substrate. Combined with the boron-supply dependence and the correlated Raman/SEM measurements, these results show that alloy-side BN formation, sapphire-side hBN formation, and development of the local hBN--sapphire architecture represent distinct but coupled aspects of the Ni--Cr-mediated growth process.

\section{Conclusions}

This work demonstrates, to our knowledge, the first successful sapphire-side hBN formation by Ni--Cr-mediated growth. hBN forms on the gas-exposed Ni--Cr alloy throughout the investigated boron-supply range, whereas sapphire-side hBN emerges only at the higher relative boron supplies of Alloy:B = 24:1 and 18:1, accompanied by additional crystalline reflections and extensive substrate reconstruction. The reconstructed outer surface remains predominantly Al/O-rich, indicating that this transformation is primarily substrate-derived.  These results show that conditions sufficient for hBN formation on the Ni--Cr alloy are not sufficient for hBN formation on sapphire. Increased boron supply drives the substrate side into a distinct reaction regime that enables hBN crystallization while simultaneously transforming the sapphire surface.  The key challenge is therefore to decouple sapphire-side hBN formation from excessive substrate reconstruction. Further control of boron supply and metal--sapphire interfacial reactions should clarify the growth mechanism and guide the development of a more uniform hBN/sapphire interface.

\section{acknowledgement}

This work was supported by the National Research Foundation of Korea (NRF) grant funded by the Korea government (MSIT) (RS-2021-NR060087, RS-2023-00240724, RS-2025-25457100, RS-2022-NR068228, RS-2026-25509069, IITP-2026-RS-2024-00437191) and through Korea Basic Science Institute (National Research Facilities and Equipment Center) grant (2021R1A6C101A437) funded by the Ministry of Education.


%

\end{document}